\documentclass[aps,prd,twocolumn,groupedaddress,amsmath,amssymb,superscriptaddress]{revtex4}
\usepackage{graphicx}
\renewcommand{\vec}[1]{\mathbf{#1}}
\begin{document}
\title{Demonstrating the Feasibility of Line Intensity Mapping Using Mock Data of Galaxy Clustering from Simulations}

\author{Eli Visbal}
\email[]{evisbal@fas.harvard.edu}
\affiliation{Jefferson Laboratory of Physics, Harvard University,
Cambridge, MA 02138} 
\affiliation{Institute for Theory $\&$ Computation, Harvard University,
60 Garden Street, Cambridge, MA 02138}

\author{Hy Trac}
\email[]{hytrac@cmu.edu}
\affiliation{ Department of Physics, Carnegie Mellon University,
Pittsburgh, PA 15213}

\author{Abraham Loeb}
\email[]{aloeb@cfa.harvard.edu}
\affiliation{Institute for Theory $\&$ Computation, Harvard University,
60 Garden Street, Cambridge, MA 02138}

\date{\today}

\begin{abstract}
Visbal $\&$ Loeb (2010) have shown that it is possible to measure the
clustering of galaxies by cross correlating the cumulative emission
from two different spectral lines which originate at the same
redshift.  Through this cross correlation, one can study
galaxies which are too faint to be individually resolved. This
technique, known as intensity mapping, is a promising probe of the
global properties of high redshift galaxies.  Here, we test the
feasibility of such measurements with synthetic data generated from
cosmological dark matter simulations.  We use a simple prescription
for associating galaxies with dark matter halos and create a
realization of emitted radiation as a function of angular position and
wavelength over a patch of the sky.  This is then used to create
synthetic data for two different hypothetical instruments, one aboard
the Space Infrared Telescope for Cosmology and Astrophysics (SPICA)
and another consisting of a pair of ground based radio telescopes
designed to measure the CO(1-0) and CO(2-1) emission lines.  We find
that the line cross power spectrum can be measured accurately from the
synthetic data with errors consistent with the analytical prediction
of Visbal $\&$ Loeb (2010).  Removal of astronomical backgrounds and
masking bright line emission from foreground contaminating galaxies do
not prevent accurate cross power spectrum measurements.


\end{abstract}

\maketitle

\section{Introduction}
Recently, Visbal $\&$ Loeb  (2010) \cite{VL} suggested a new technique for statistically observing the clustering of faint
galaxies through intensity mapping of multiple atomic and molecular lines (see also \cite{asantha_IM,carilli_IM,Lidzetal}).  
This method can probe galaxies which are too faint
to be seen individually, but which contribute significantly to the
cumulative emission due to their large numbers.

Atoms and molecules in the interstellar medium of galaxies produce
line emission at particular rest frame wavelengths \cite{binney}.  For
galaxies at cosmological distances, these wavelengths are redshifted
by a factor of $(1+z)$ due to cosmic expansion.  Thus,
for emission in a particular spectral line, the observed angular
position and the observed wavelength correspond to a 3D spatial location.
With observational data which includes both spectral
and spatial information, one can then measure the three dimensional
clustering of galaxies.

Before line emission can be associated with a particular location in
space, one must separate it from spectrally extended emission.
Galactic continuum emission and spectrally smooth astrophysical
foregrounds and backgrounds (e.g., the Cosmic Microwave Background or
galactic dust emission) can be removed by fitting smooth functions of
frequency to data and subtracting them away; this has been discussed
extensively in the context of cosmological 21cm observations
\cite{2010arXiv1010.4109P,2005ApJ...619..678M,Wang:2005zj,2006ApJ...653..815M,2009MNRAS.398..401L,liutegmark}.
After background emission is removed one still needs to avoid possible
confusion with other emission lines.  For multiple lines of different
rest frame wavelengths the intensity at a particular observed
wavelength corresponds to emission from multiple redshifts, one for
each emission line.  With both spatial and spectral information, the
total emission over a small range in observed wavelength corresponds
to a superposition of the 3D distribution of galaxies at different
redshifts.

\begin{figure*}
\includegraphics[width=2.5in,height=2.5in]{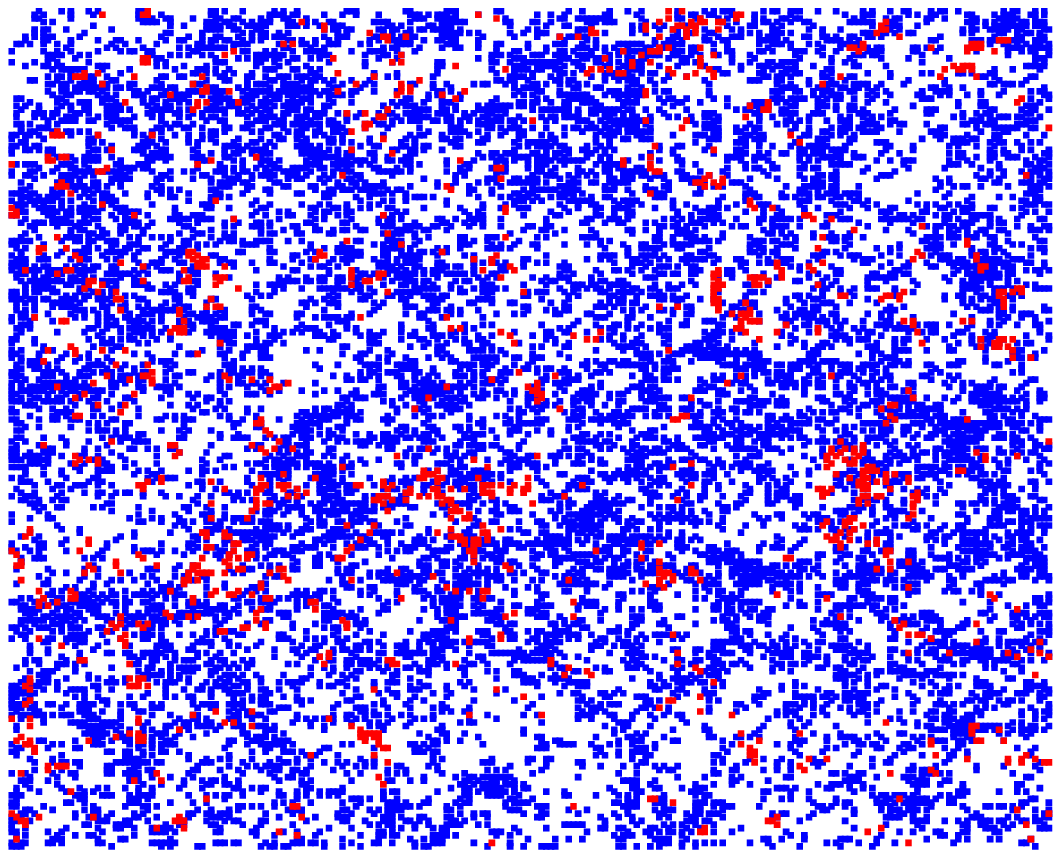}
\includegraphics[width=2.5in,height=2.5in]{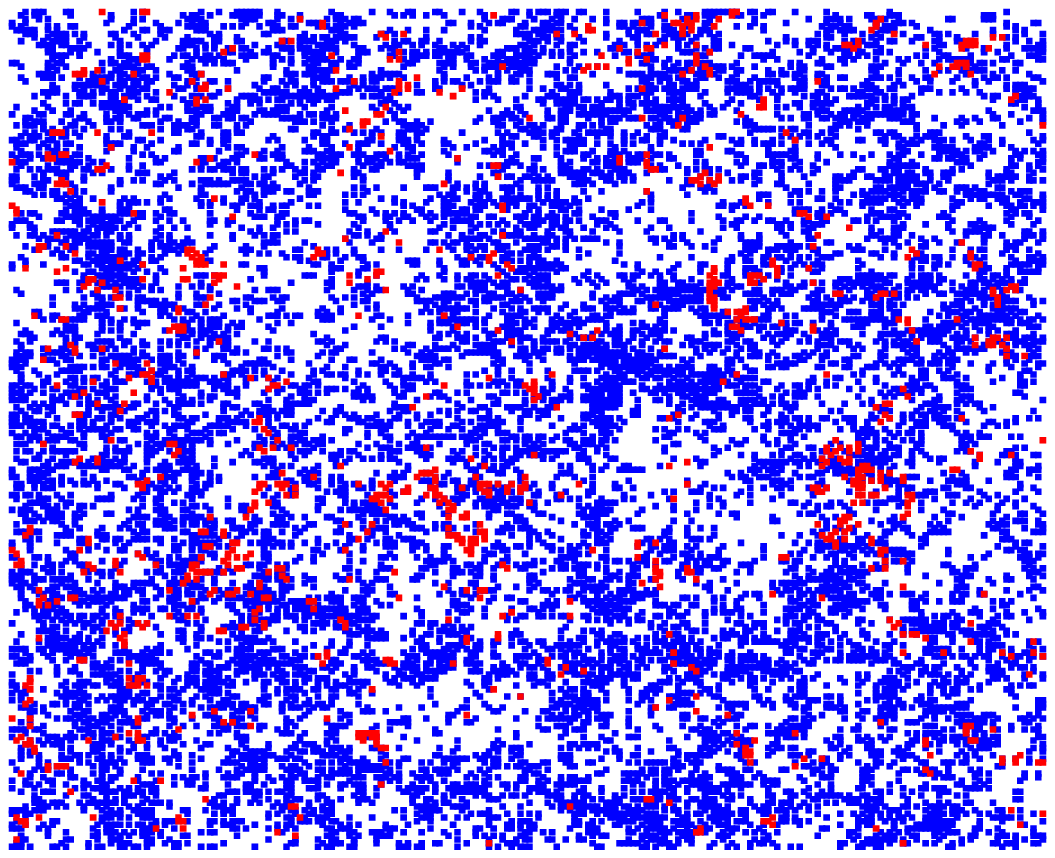}
\caption{\label{image} A slice from our simulated realization of line
emission from galaxies at an observed wavelength of 441$\mu$m (left)
and 364$\mu$m (right).  The slice is in the plane of the sky and spans
$250\times250$ comoving Mpc$^2$ with a depth of $\Delta \nu / \nu =
0.001$. The colored squares indicate pixels in our SPICA example
(presented below) which have line emission greater than $200 {\rm
Jy/Sr}$ for the left panel and $250 {\rm Jy/Sr} $ for the right panel.
The emission from OI(63$\mu$m) and OIII(52$\mu$m) is shown in red on
the left and right panels, respectively, originating from the same
galaxies at $z=6$.  All of the other lines in Table \ref{tab} are
included and plotted in blue.  Cross correlating data at these two
observed wavelengths would reveal the emission in OI and OIII from
$z=6$ with the other emission lines being essentially uncorrelated. }
\end{figure*}

Fortunately, it is possible to statistically isolate the fluctuations
from a particular redshift by cross correlating the emission in two
different lines \cite{VL}.  If one compares the fluctuations at two different
wavelengths, which correspond to the same redshift for two different
emission lines, the fluctuations will be strongly correlated.
However, the signal from any other lines arises from galaxies at
different redshifts which are very far apart and thus will have much
weaker correlation (see Figure \ref{image}).  In this way, one can
measure either the two-point correlation function or power spectrum of
galaxies at some target redshift weighted by the total emission in the
spectral lines being cross correlated.

We emphasize that one can measure the line cross power spectrum from
galaxies which are too faint to be seen individually over detector
noise.  Hence, a measurement of the line cross power spectrum can
provide information about the total line emission from all of the
galaxies which are too faint to be directly detected.  One possible
application of this technique would be to measure the evolution of
line emission over cosmic time to better understand galaxy evolution
and the sources that reionized the Universe.  Changes in the minimum
mass of galaxies due to photoionization heating of the intergalactic
medium during reionization could also potentially be measured
\cite{VL}.

Here we use cosmological simulations to test the feasibility of
measuring the galaxy line cross power spectrum.  We create synthetic
data sets for two hypothetical instruments, one on the Space Infrared
Telescope for Cosmology an Astrophysics (SPICA) and the other
consisting of a pair of ground based radio telescopes optimized to
measure CO(1-0) and CO(2-1) emission from high redshifts.  We test how
well the cross power spectrum can be measured and find agreement with
the analytical expectation derived in \cite{VL}.  However there are
some additional complications.  Small $k$-modes along the line of sight
which are contaminated during the foreground removal process must be
discarded, increasing the statistical uncertainty on large spatial
scales.  Additionally, when masking out contaminating emission lines
from bright foreground galaxies one must be careful not to introduce a
spurious correlation between the data sets being cross correlated.

The paper is organized as follows.  In $\S$2 we describe the methods
used in this paper.  This includes a brief review of the galaxy line
cross power spectrum, a description of the synthetic data sets, the
details of the simulations, and a discussion of the steps involved in
measuring the cross power spectrum.  In $\S$3 and $\S$4 we present our
results for the SPICA example and the CO(1-0) and CO(2-1) telescopes,
respectively.  Finally, we discuss and summarize our conclusions in
$\S$5.  Throughout, we assume a $\Lambda$CDM cosmology with
$\Omega_\Lambda=0.73$, $\Omega_m=0.27$, $\Omega_b=0.045$, $h=0.7$,
$n_{\rm s}=0.96$ and $\sigma_8=0.8$, \cite{wmap7}.

\section{Method}
\subsection{Galaxy line cross power spectrum}
First, we briefly review the galaxy line cross power spectrum.  For a
more complete discussion, see Visbal $\&$  Loeb (2010) \cite{VL}.  We assume that emission is
measured both as a function of angle on the sky and observed
wavelength.  If one fits a smooth function of wavelength along each
direction on the sky and subtracts it from the data, one obtains the
fluctuations from the average signal as a function of angle and
wavelength: $\Delta
S(\theta_1,\theta_2,\nu)=S(\theta_1,\theta_2,\nu)-\bar{S}$.  There is
a one to one correspondence between angular position and wavelength
and spatial position for emission in a particular line.
For convenience we use comoving coordinates at the location of
the target galaxies instead of angle and wavelength.  The fluctuations
at a particular location results from a number of different sources,
\begin{equation}
\label{deltas}
 \Delta S_1=\Delta S_{\rm line1}+\Delta S_{\rm noise}+\Delta S_{\rm badline1}+\Delta S_{\rm badline2} + \ldots
\end{equation}
which include contributions from the target galaxies we wish to cross
correlate, detector noise, and emission in different lines from
galaxies at different redshifts which we refer to as ``bad line''
emission.  One can cross correlate the fluctuations in two different
lines from the same galaxies.  We define the line cross correlation
function as,
\begin{equation}
\xi_{1,2}(\vec{r})=\langle \Delta S_{1}(\vec{r}_o,\vec{x}) \Delta S_{2}(\vec{r}_o,\vec{r}+\vec{x}) \rangle,
\end{equation}
where subscripts denote different lines being cross correlated.  The
center of the survey volume is denoted by $\vec{r}_o$, $\vec{x}$ is
the distance from the center in the first set of fluctuations, and
$\vec{r}+\vec{x}$ is the distance from the center in the second set of
fluctuations.

Because the noise fluctuations in the two different data sets are
uncorrelated and galaxies seen in different bad lines will have very
large separations and thus be essentially uncorrelated we are only
left with contributions from the target galaxies.  On large scales we
can make the assumption that line fluctuations due to galaxy
clustering are given by $\Delta S_{\rm line1}=\bar{S}_1 \bar{b}\delta(\vec{r})$,
where $\bar{S}_1$ is the average target line signal, $\bar{b}$ is the
luminosity weighted average galaxy bias, and $\delta(\vec{r})$ is the
cosmological over-density at a location $\vec{r}$.  It follows that,
\begin{multline}
\xi_{1,2}(\vec{r})=\langle \Delta S_{\rm line1}(\vec{r}_o,\vec{x}) \Delta S_{\rm line2}(\vec{r}_o,\vec{r}+\vec{x}) \rangle \\ =\bar{S}_1 \bar{S}_2 \bar{b}^2 \langle \delta(\vec{x}) \delta(\vec{r}+\vec{x}) \rangle=\bar{S}_1 \bar{S}_2 \bar{b}^2 \xi(\vec{r}),
\end{multline}
where $\xi(\vec{r})$ is the cosmological matter correlation function and
the subscript numbers denote the different lines being cross correlated.

The line cross power spectrum is then defined as the Fourier
transform,
\begin{equation}
P_{1,2}(\vec{k}) = \int d^3\vec{r} \xi_{1,2}(\vec{r}) e^{i \vec{k} \cdot \vec{r}}=\bar{S}_{1}\bar{S}_{2}\bar{b}^2 P(\vec{k}) + P_{\rm shot},
\end{equation}
where $P_{\rm shot}$ is the shot-noise power spectrum due to the
discrete nature of galaxies.

An unbiased estimator for the cross power spectrum is given by the
product of the Fourier transforms of the data sets,
\begin{equation}
\label{estimator}
\hat{P}_{1,2}=\frac{V}{2} (f^{(1)}_{\vec{k}}f^{(2)*}_{\vec{k}}+f^{(1)*}_{\vec{k}}f^{(2)}_{\vec{k}} ),
\end{equation}
where $V$ is the volume of the survey and the superscripts denote the
different lines being cross correlated.  The Fourier amplitude is given by,
\begin{equation}
\label{famp}
f_{\vec{k}} = \int d^3\vec{r} \Delta S(\vec{r_o},\vec{r})W(\vec{r})e^{i\vec{k} \cdot \vec{r}}.
\end{equation}
Here $W(\vec{r})$ is a window function that is constant over
the survey volume and zero at all other locations.  It is
normalized such that, $\int W(\vec{r}) d^3\vec{r}=1$. 

The root mean square (RMS) error in a measurement of the cross power spectrum
at one particular $k$-value is given by \cite{VL},
\begin{equation}
\label{ccerror}
\delta P^2_{1,2}=\frac{1}{2}(P_{1,2}^2 + P_{\rm1total}P_{\rm2total}),
\end{equation}
where $P_{\rm1total}$ and $P_{\rm2total}$ are the total power spectrum
corresponding to the first line and second line being cross
correlated.  Each of these includes a term for the power spectrum for
each of the bad lines, the target line, and detector noise (see
Appendix A of Ref.~\cite{VL}).  When averaging nearby values of the power
spectrum this error goes down by a factor of $\sqrt{N_{\rm modes}}$,
where $N_{\rm modes}$ is the number of statistically independent
$k$-values at which the power spectrum is measured.


\subsection{Synthetic data set}
In order to test the feasibility of measuring the line cross power
spectrum we create synthetic data sets for instruments measuring both
spatial and spectral information.  Our goal is to produce a
realization of the light from all galaxies as a function of angular
position and observed wavelength on a patch of the sky.  We create
these data with a cosmological dark matter simulation (described in
detail below).  From the simulation we construct a light cone which
has the distribution of dark matter halos which would be observed
today in the volume corresponding to an angular patch on the sky out
to a redshift of $z=10$.

A simple prescription is used to associate galaxies with the dark
matter halos from our simulation.  We assign each galaxy a spectrum
and assume that its intensity scales with star formation rate (SFR).
The SFR versus halo mass relation is determined by
matching comoving density with observed UV luminosity functions
\cite{LF1,LF2,LF3,LF4,LF5}.

We assume that galaxies are found in dark matter halos above a minimum
mass, $M_{\rm min}$.  After reionization $M_{\rm min}$ represents the
threshold for assembling heated gas out of the photo-ionized
intergalactic medium, corresponding to a minimum virial temperature of
$\approx 10^5\rm{K}$ \cite{Wyithe:2006st}.  We assume that reionization was
completed by a redshift of $z \approx 10$.  In all of the examples
presented below, $M_{\rm min}$ is set to correspond to this
post-reionization requirement.

The larger dark matter halos in our simulation may host multiple
galaxies.  To incorporate this effect in our synthetic data we have
used a simple prescription for the halo occupation distribution.
Following Ref.~\cite{2004ApJ...609...35K} for the distribution
of dark matter sub-halos, we consider two different types of galaxies:
central and satellite.  We assume that the distribution of central
galaxies is a step function: above $M_{\rm min}$ we assume each halo has one galaxy at
its center.  We then assume that there are a number of satellite
galaxies given by a Poisson distribution with a mean of $N_{\rm
sat}=\left ( M/M_1 \right )^\beta$, here $\beta=1$, and $M_1=30M_{\rm
min}$ at $z=0-0.5$; $M_1=20M_{\rm min}$ at $z=0.5-2$; and
$M_1=10M_{\rm min}$ at $z>2$.  We distribute these galaxies randomly,
but weighted by an NFW profile, throughout the larger host dark matter
halo.  We treat the central and satellite galaxies as independent in assigning 
 star formation rates to them as explained below.  We associate
half of the total halo mass to the central galaxy halos and split the
remainder of the mass equally to all of the satellite galaxy halos.

After relating galaxies to dark matter halos in the simulation we
produce a spectrum for each galaxy.  For
the continuum, we take the measured spectral energy distribution of
M82 and scale it with the SFR \cite{M82cont}.
The results are insensitive to the particular choice of galaxy continuum, as it
is removed in the fitting and subtraction stage of the data analysis, as
discussed below.

In order to estimate the amplitude of line emission fluctuations we
assume a linear relationship between line luminosity, $L$, and star
formation rate, $\dot{M}_*$, $L=\dot{M}_* \times R$, where $R$ is the
ratio between SFR and line luminosity for a particular
line.  This is similar to existing relations in different bands (see
Ref.~\cite{Kennicutt}) and was used in the past to estimate the
strength of the galactic lines we consider \cite{2008A&A...489..489R}.
The values for relevant lines are shown in Table \ref{tab}.  For the
first 7 lines, we use the same ratios, $R$, as in
Ref.~\cite{2008A&A...489..489R} which were calculated by taking the
geometric average of the ratios from an observational sample of lower
redshift galaxies \cite{2001ApJ...561..766M}.  The other lines have
been calibrated based on the galaxy M82 \cite{M82lines}.  We assign a
width to the lines based on the circular virial velocity of the dark
matter halos, but the results are mostly insensitive to this choice
for the spectral resolutions we consider in our examples.  This is
because the majority of the signal comes from lines which are
spectrally unresolved.

For the results presented below we have made the simplification that
all galaxies have the same R value for each emission line.  Even if there is random scatter
in the R values in each galaxy, the line cross
power spectrum will remain unchanged.  This scatter will behave
essentially like detector noise with intensity that is non-uniform
across the data cube.

\begin{table}
\caption{\label{tab}Ratio between line luminosity, $L$, and star formation rate,
$\dot{M}_*$, for various lines.  For the
first 7 lines this ratio is measured from a sample of low redshift
galaxies. The other lines have been calibrated based on the galaxy M82.
}
\begin{tabular}{c c c}
\hline
Species & Emission Wavelength[$\mu$m] & R[$L_{\odot}/(M_{\odot}/{\rm yr})$] \\
\hline
CII & 158 & $6.0 \times 10^6$\\
OI  & 145 & $3.3 \times 10^5$\\
NII & 122 & $7.9 \times 10^5$\\
OIII & 88 & $2.3 \times 10^6$\\
OI & 63 & $3.8 \times 10^6$\\
NIII & 57 & $2.4 \times 10^6$\\
OIII & 52 & $3.0 \times 10^6$\\
$^{12}$CO(1-0)  &2610 & $ 3.7 \times 10^3$\\
$^{12}$CO(2-1)  &1300 & $ 2.8 \times 10^4$\\
$^{12}$CO(3-2)  &866 & $  7.0 \times 10^4$\\
$^{12}$CO(4-3)  &651 & $  9.7 \times 10^4$\\
$^{12}$CO(5-4)  &521 & $  9.6 \times 10^4$\\
$^{12}$CO(6-5)  &434 & $  9.5 \times 10^4$\\
$^{12}$CO(7-6)  &372 & $  8.9 \times 10^4$\\
$^{12}$CO(8-7)  &325 & $  7.7 \times 10^4$\\
$^{12}$CO(9-8)  &289 & $  6.9 \times 10^4$\\
$^{12}$CO(10-9) &260 & $  5.3 \times 10^4$\\
$^{12}$CO(11-10) &237  & $ 3.8 \times 10^4$\\
$^{12}$CO(12-11) &217  & $ 2.6 \times 10^4$\\
$^{12}$CO(13-12) &200  & $ 1.4 \times 10^4$\\
CI  & 610 & $ 1.4 \times 10^4$\\
CI  & 371 & $ 4.8 \times 10^4$\\
NII & 205 & $ 2.5    \times 10^5$\\
$^{13}$CO(5-4) & 544 & 3900\\
$^{13}$CO(7-6) & 389 & 3200 \\
$^{13}$CO(8-7) & 340 & 2700 \\
HCN(6-5) & 564 & 2100 \\
\hline
\end{tabular}
\end{table}

We use observed UV luminosity functions (\cite{LF1,LF2,LF3,LF4,LF5}) of
galaxies to calibrate the SFR assigned to dark matter halos with an
abundance matching technique.  Given the observed luminosity functions, we
determine the number density of galaxies as a function of SFR through the
relation,
\begin{equation}
L_{\rm{UV}} = L_\lambda \left ( \frac{\dot{M}_*}{M_\odot \rm{yr}^{-1}} \right ){\rm ergs/s/Hz},
\end{equation}
where $L_\lambda$ is given by $L_\lambda = 8 \times 10^{27}$ at a
rest frame wavelength of $\lambda = 1500{\rm \AA}$.  This assumes a
Salpeter initial mass function from $0.1-125 M_{\odot}$ and a constant
 $\dot{M}_* \gtrsim 100 {\rm Myr}$.  The relationship between halo mass,
$M_i$, and SFR $\dot{M}_{*i}$ at some particular mass,
is found from the relation $n_h(>M_{i})=n_g(>\dot{M}_{*i})$.  Here
$n_h(>M)$ is the number density of dark matter halos above mass $M$ in
our simulation and $n_g( >\dot{M}_*)$ is the number density of galaxies
implied by the UV luminosity function above the SFR
value, $\dot{M}_{*}$.  This procedure is carried out in a number of different
redshift bins which cover our entire light cone.  As a simple
correction for attenuation due to dust we increase the SFR of all
halos in each redshift bin by a factor which sets the global SFR equal
to that given in Ref.~\cite{LF3} (the blue solid curve in Figure 10 of
Ref.~\cite{LF3}).  In the highest redshift bin we do not apply any
dust correction.  The particular parameters used for the abundance
matching procedure are listed in Table \ref{lumtable}.

\begin{table}
\caption{\label{lumtable} Schechter function
parameters for the UV Luminosity Functions used to assign SFR to dark matter
halos.  These parameters are used through abundance matching.}
\begin{tabular}{c c c c c}
\hline
$z$ & $\phi^*(\times 10^{-3} \rm{Mpc^{-3}})$  & $M^*_{\rm AB}$ & $\alpha$ & Ref.  \\
\hline
0.0-0.5  & 4.07 & -18.05 & -1.21 &  \cite{LF1}\\
0.5-1.0 & 3.0 & -19.17 & -1.52  &  \cite{LF2}\\
1.0-1.5 & 1.26 & -20.08 & -1.84 &  \cite{LF2}\\
1.5-2.0 & 2.3 & -20.17 & -1.60  &  \cite{LF2}\\
2.0-2.7 & 2.75 & -20.7 & -1.73  &  \cite{LF3}\\
2.7-3.4 & 1.71 & -20.97 & -1.73 &  \cite{LF3}\\
3.4-4.5 & 1.3 & -20.98 & -1.73  &  \cite{LF4}\\
4.5-5.5 & 1.0 & -20.64 & -1.66  &  \cite{LF4}\\
5.5-6.5 & 1.4 & -20.24 & -1.74  &  \cite{LF4}\\
6.5-10.5 & 0.86 & -20.14 & -2.01  &  \cite{LF5}\\          
\hline
\end{tabular}
\end{table}

\begin{figure*}
\includegraphics[width=5in]{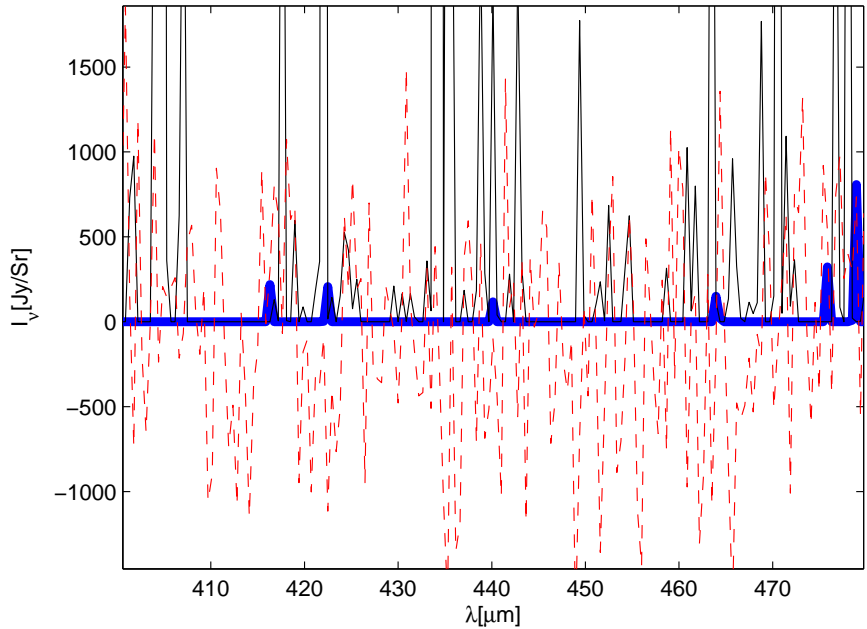}
\caption{\label{LOSplot} Various components of the synthetic data set
 for a typical line of sight.  We plot data from the SPICA example
 discussed below.  The thick blue curve is the line emission from the
 target galaxies, the thin red dashed curve is the contribution from
 detector noise, and the thin black curve is the emission from all of
 the bad lines.  We have not included the bright astrophysical
 foregrounds because they are orders of magnitude greater than all of
 the components plotted here.  This emission along with galaxy
 continuum (not plotted) is removed in the fitting and subtraction
 step of measuring the power spectrum discussed in the text. }
\end{figure*}

Finally, we add detector noise and bright astronomical foreground
and background emission.  For the examples below we include both the
CMB and emission from dust in our galaxy.  The dust emission is
treated as a black body with a $\nu^2$ emissivity scaled to match the
background radiation measured by COBE FIRAS in the faintest area on
the sky \cite{FIRAS}.  In Figure \ref{LOSplot}, we illustrate the
different components which make up our data sets.

\subsection{Simulations} 
To create our synthetic data we simulate the light cone of dark matter
in a 100$\times$100 arcmin$^2$ angular patch of the sky out to high
redshift.  We use a particle-multi-mesh N-body code to evolve the dark
matter distribution \cite{2006NewA...11..273T}. The simulation outputs
are then stacked along the line of sight out to z = 10.

For most of our light cone, we use an N-body simulation with
2048$^3$ dark matter particles on an effective mesh with 7680$^3$
cells in a comoving box with a length of 200$h^{-1}{\rm Mpc}$ on a side.
This length is sufficient to cover the field of view out to the
highest redshifts of interest.  For low redshifts ($z<1$), we use a
second larger simulation to improve the sample variance of large
halos.  This simulation also contains 2048$^3$ dark matter particles
and a mesh of 7680$^3$ cells, but has a length of 400$h^{-1}{\rm Mpc}$ on
each side of the box.  In both simulations we identify dark matter
halos using a spherical overdensity algorithm.  This is done by
examining snapshots taken every 20 Myr and 40 Myr in the
200$h^{-1}{\rm Mpc}$ and 400$h^{-1}{\rm Mpc}$ simulations respectively.

The light cone is constructed from a series of redshift zones, each
 zone spanning one comoving box length. Each zone is constructed from
 several redshift shells of thickness corresponding to a time interval
 of 20 or 40 Myr depending on the box size. The shells are stacked in
 a continuous fashion, but the zones are randomized to eliminate any
 very long artificial structure.  This produces a discontinuity across
 zone boundaries.  We are careful to only measure the power spectrum
 within one zone for our examples to avoid any problems associated
 with this discontinuity.

\begin{figure*}
\includegraphics[width=5in,height=5in]{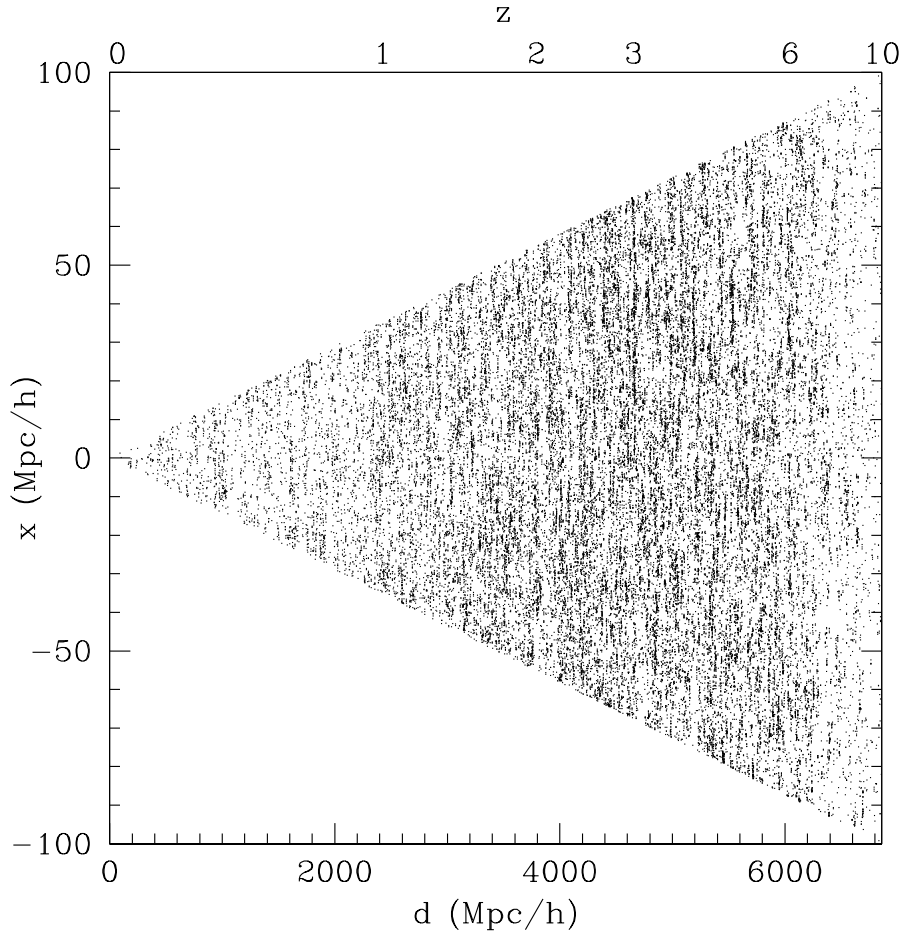}
\caption{\label{complot} The positions of dark matter halos from a
slice of our simulated light cone projected onto the x-z plane (where
z is the direction along the line of sight).  The slice has thickness
$\Delta y=0.5 h^{-1} {\rm Mpc}$.  Each dark point represents a dark matter
halo.  Our light cone corresponds to $100\times100$ arcmin$^2$ on the
sky.  The details are explained in $\S$2.}
\end{figure*}

\subsection{Cross power spectrum measurement}
Measuring the cross power spectrum consists of three main steps:

\noindent 1. Fitting a smooth function of wavelength to each pixel and
   subtracting it away (note that we term each line of sight on the
   sky a ``{\bf \emph{pixel}}'' and each spectral component of the
   3D data cube a ``{\bf \emph{voxel}}'').

\noindent 2. Masking out \emph{voxels} with bright bad line emission.

\noindent 3. Taking the product of the Fourier modes to estimate the power
spectrum and then averaging in spherical shells in $k$-space.

We discuss these in turn.  The fitting stage is necessary because we
 seek to measure only the signal coming from line emission and our
 data contains signal from both galaxy continuum emission as well as
 bright astrophysical foregrounds and backgrounds.  Since these other
 sources vary slowly in the spectral direction we can remove them by
 fitting a smooth function of wavelength to each \emph{pixel} on the sky and
 subtracting it.  This is the same procedure which has been discussed
 extensively in the context of cosmological measurements of 21cm
 radiation from neutral hydrogen
 \cite{2010arXiv1010.4109P,2005ApJ...619..678M,Wang:2005zj,2006ApJ...653..815M,2009MNRAS.398..401L}.

More specifically, with our data sets we fit a polynomial in
wavelength to the spectrum in each \emph{pixel} and then subtract it away.
This removes the foregrounds and galaxy continuum, as well as some
large scale fluctuations in line emission along the line of sight.  In
order to minimize loss of the line signal we do not include \emph{voxels} in
our fit which contain bright line emission.  We do this by an iterative fit: 
we fit once to remove the foregrounds and identify the bright
\emph{voxels} and then fit again excluding them.

There will necessarily be some signal lost on large scales as a result
of the fitting and subtraction stage.  Fortunately as discussed in
Ref.~\cite{2010arXiv1010.4109P}, if we decompose our signal into Fourier
modes, the lost signal is only from small $k$-modes (corresponding to
long wavelengths) along the line of sight.  If we exclude these
corrupted $k$-modes in step 3 of measuring the cross power spectrum, we still have an unbiased estimation of the cross
power spectrum without subtraction losses.  Note that throwing away
the low $k$-modes does have a price.  Since there are fewer statistical
samples of modes this procedure increases the variance of power
spectrum measurements on large scales.  Because we wish to minimize
the number of these corrupted modes, we fit with the lowest order
polynomial which leaves no significant residual foregrounds.

After we have subtracted away the foreground and continuum emission it
is necessary to remove \emph{voxels} with very bright bad line emission.
This is necessary because even though line emission from bright
foreground galaxies does not bias our measurements of the power
spectrum it does increase the error of our measurements due to
the contribution in Eq.~(\ref{ccerror}).

The masking procedure must be done carefully in order to not introduce
spurious correlations between the two data sets being cross
correlated.  For example, if one simply sets all \emph{voxels} above some
threshold signal equal to zero, a spurious change to the cross power
spectrum is introduced (see Fig. \ref{badmask}).  This is because the
location of the brightest \emph{voxels} (mainly due to contaminating bright
foreground galaxies) are correlated with the distribution of target
line emission.  The signal from the bad lines and the target lines
overlap so that bright bad lines which appear in the data at locations
of over-densities in the target lines are more likely to be above the
removal threshold.  Thus, the bad lines left after masking in one data
cube will be anti-correlated with the target lines in the other cube.
This causes the measured cross power spectrum to be lower than what
would be measured from the target lines alone.

In order to avoid this type of complication one can mask out \emph{voxels} in
a way which is uncorrelated with the target line emission being
measured.  This can be done by identifying individual bright sources
instead of just removing the brightest \emph{voxels} in the data.  The \emph{voxels}
with bright contaminating lines can then be set to zero.  These
sources could be identified by looking at a series of different
wavelengths and identifying them with multiple lines.  Entirely different
surveys could also be used to determine where contaminating lines from
bright foreground galaxies will appear and be removed.  In our
examples below, we assume that all of the galaxies which emit lines
brighter than five times the RMS detector noise can be identified
directly.  When setting the masked \emph{voxels} to zero we treat this as a
change in the window function, $W(\vec{r})$, which appears in
Eq.~(\ref{famp}).  We normalize this new window function such that $\int
W(\vec{r}) d^3\vec{r}=1$. 

In the final step, we take the discrete Fourier transform of the two
3D data cubes being cross correlated.  The estimation of the power
spectrum at some particular $k$ value is then given by the real part
of the product of the survey volume, the Fourier mode of one data set,
and the complex conjugate of the same Fourier mode in the other data
set.  This is equivalent to Eq.~(\ref{estimator}).  Finally, we break
$k$-space into spherical shells with uniform thickness in $\log(k)$.
We then take the average estimated power spectrum of all the modes
contained within each shell.  As discussed above, we do not include
low $k$-modes along the line of sight which have been contaminated
during the fitting and subtraction stage.  Specifically, we do not
include $k$-modes which have a component along the line of sight
smaller than, $k_{\rm cut}$, the lowest value for which there is no
significant contamination.  In the examples below we find that for
$k_{\rm cut}=0.06 h{\rm Mpc^{-1}}$ there is no significant loss
of power due to the foreground removal process.

\section{SPICA} 
\subsection{Instrument}
We consider two different examples of instruments and lines which
could be used to measure the galaxy line cross power spectrum.  In our
first example, we envision an instrument on the planned Space Infrared
Telescope for Cosmology and Astrophysics (SPICA) \cite{SPICAref}.
SPICA is a 3.5 meter space-borne infrared telescope planned for launch
in 2017.  It will be cooled below $5$K, providing measurements which
are orders of magnitude more sensitive than those from current
instruments.  We consider an instrument based on the proposed high
performance spectrometer $\mu$-spec (H. Moseley, private communication
2009).  This instrument will provide background limited sensitivity
with wavelength coverage from $~250-700 \mu m$.  A number of
$\mu$-spec units will be combined to record both spatial and spectral
data in each pointing, which will be perfectly suited for intensity
mapping.  We assume that spectra for $100$ diffraction limited beams
can be measured simultaneously with a resolving power of
$R=(\nu/\Delta \nu)=1000$.

\begin{figure*}
\includegraphics[width=5in]{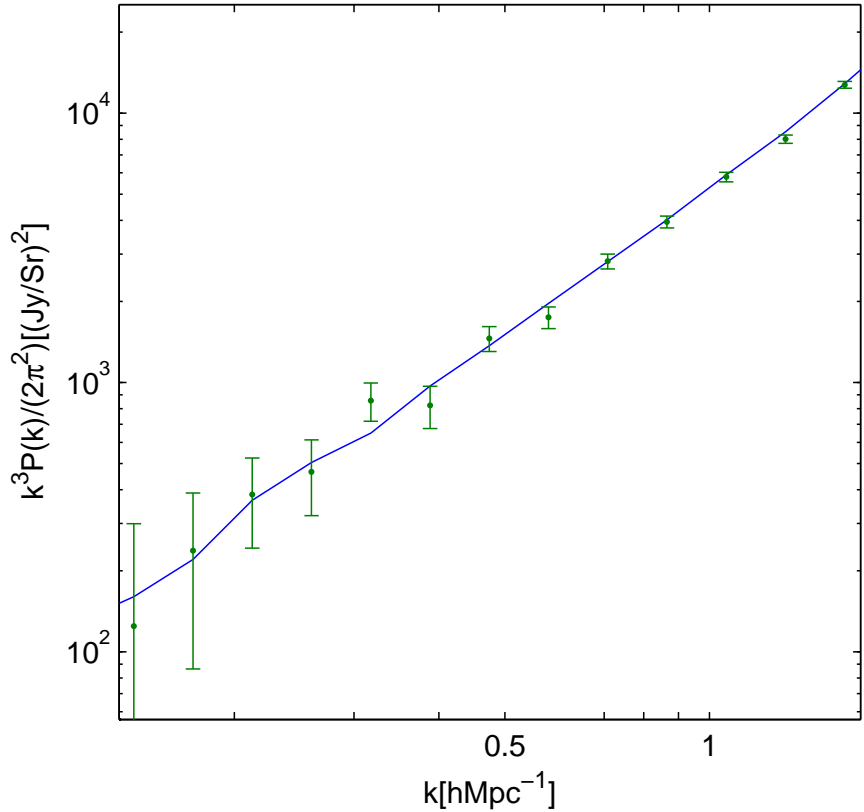}
\caption{\label{spicaplot} The cross power spectrum of OI(63 $\mu$m)
and OIII(52 $\mu$m) at $z=6$ measured from simulated data for our
hypothetical instrument modeled after SPICA.  The blue curve is the
cross power spectrum measured when only line emission from galaxies in
the target lines is included.  The green points are the recovered
power spectrum when detector noise, bad line emission, galaxy
continuum emission, and bright astrophysical foreground and background
emission (i.e. dust in our galaxy and the CMB) are included.  The
error bars are the theoretical prediction of the root mean square
error derived in \cite{VL} and given by Eq.~(\ref{ccerror}).  In
determining the error bars we have estimated $P_{\rm1total}$ and
$P_{\rm2total}$ using our simulated data.  These errors include detector
noise, bad line emission and sample variance.}
\end{figure*}

\begin{table*}
\caption{\label{Otable} Summary of the results from the example of
cross correlating OI(63 $\mu$m) and OIII(52 $\mu$m) with SPICA.  The
RMS detector noise is the value in each \emph{voxel}.  The bad line
power to detector noise power ratio gives the relative contributions to
the statistical error in the cross power spectrum due to the
auto-correlations from all the bad lines and the detector noise which
appear in Eq.~(\ref{ccerror}).  }
\begin{tabular}{l l l}
 & OI(63 $\mu$m)  & OIII(52 $\mu$m) \\
\hline 
Average Line Signals ($\bar{S}_{\rm line}$) &  $20 {\rm Jy/Sr}$ & $14 {\rm Jy/Sr}$ \\ 
Fraction of Voxels Masked & 0.097 &  0.11 \\ 
RMS Detector Noise & $700{\rm Jy/Sr}$ & $400{\rm Jy/Sr}$\\
Brightness of CMB$+$Dust & $4{\rm MJy/Sr}$ & $2{\rm MJy/Sr}$ \\
Bad line Power$/$Noise Power ($k=0.3h^{-1}$Mpc) & 6.5  & 8.1  \\ 
Bad line Power$/$Noise Power ($k=1h^{-1}$Mpc) & 1.4  & 1.7  \\ 
Cross Power S$/$N per $k$-mode ($k=0.3h^{-1}$Mpc) &$\; \; \; \; \;$ 0.17    \\ 
Cross Power S$/$N per $k$-mode ($k=1h^{-1}$Mpc) & $\; \; \; \; \;$  0.14  \\ 
 \hline
\end{tabular}
\end{table*}

\begin{figure*}
\includegraphics[width=5in]{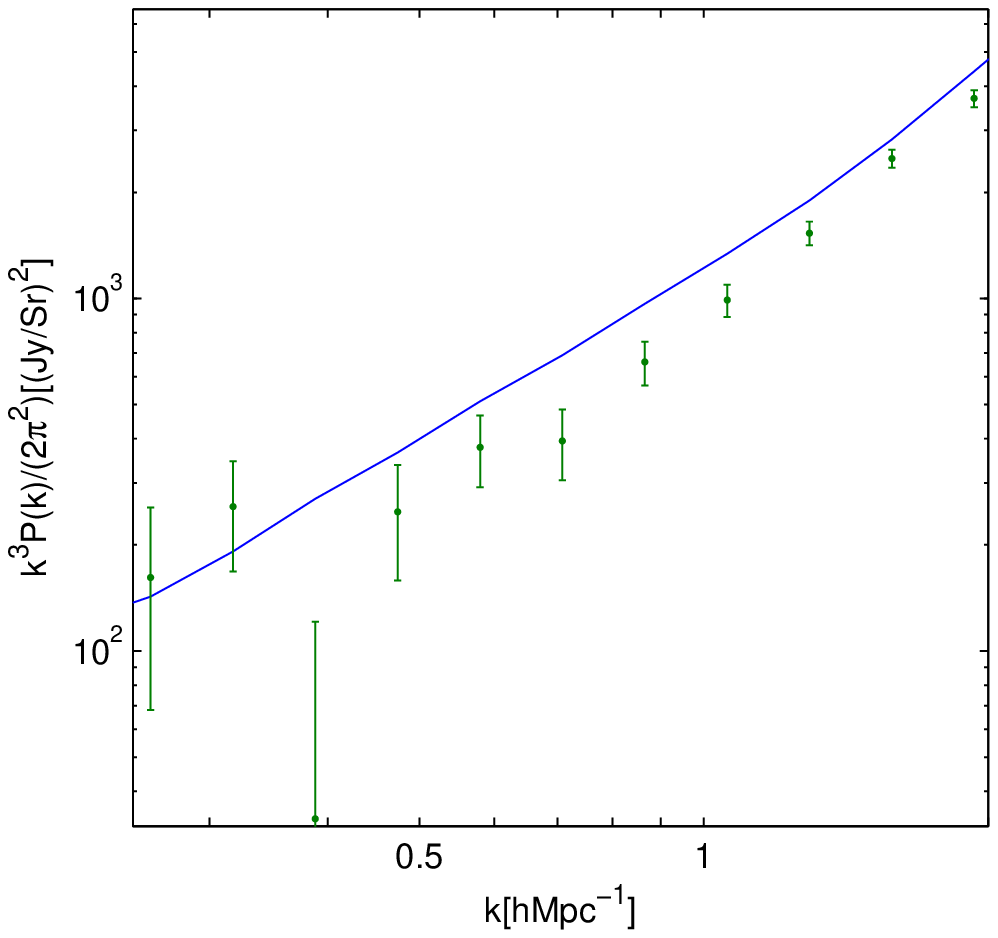}
\caption{\label{badmask} The measured power spectrum when masking is
done by simply setting the fluctuations in all \emph{voxels} with
signal greater than five times the RMS detector noise to zero.  We use
the same instrumental and survey parameters as in Figure
\ref{spicaplot} except we set the detector noise in each data cube to
zero to more clearly demonstrate the masking effect.  The points
plotted are the measurements of the cross power spectrum after all
bright \emph{voxels} have been masked and set to zero.  The error bars
show the standard deviation in the cross power spectrum of all the
modes sampled at each $k$-value.  The line is the power spectrum if
only the target galaxy lines are included (but using the same mask).
Clearly the anti-correlation between the masked bad lines and the
target lines in opposite cubes produces a systematic shift in the
power spectrum as described in the text.}
\end{figure*}

\subsection{Results}
We use the simulation described above to create a synthetic data set
and measure the cross power spectrum with the SPICA/$\mu$-spec
instrument.  We cross correlate OI(63 $\mu$m) and OIII(52 $\mu$m) from
galaxies at a redshift of $z=6$.  We assume the data covers a square
on the sky which is 1.7 degrees across (corresponding to 250 Mpc) and
a redshift range of $\Delta z =0.6$ (corresponding 280 Mpc).  We
assume a total integration time of $2 \times 10^6$ seconds spread
uniformly across this survey area.

In Figure \ref{spicaplot}, we show that using the procedure described
above we can accurately measure the cross power spectrum.  We show
both the cross power spectrum of the emission from the target lines
alone as well as that which is recovered when bad lines, detector
noise, and foregrounds are included.  The error in measuring the power
spectrum is consistent with the analytical prediction derived in
Ref.~\cite{VL}.  The details introduced in our simulation and measurements,
such as removing the foregrounds and masking out bright foreground
galaxies, does not bias our estimate of the power spectrum or increase
the uncertainty implied by Eq.~(\ref{ccerror}).  Other details of this
example are presented in Table \ref{Otable}.

In Figure \ref{badmask}, we show the effects on the measured cross
power spectrum of masking out all bright \emph{voxels}.  We have plotted the
power spectrum from the target lines alone and also with the bad lines
using the same mask in both cases.  Clearly, the anti-correlation
between the masked bad lines and the target lines in the other data
set described above has biased the cross power spectrum measurement.

We find that increasing the sky coverage (i.e. shorter integrations
for each pointing on the sky, but larger sky coverage) increases our errors in the
power spectrum.  This is due to our assumptions about masking bright
bad lines.  As the survey becomes wider the detector noise goes up and the
increased number of bright bad lines which are not masked increases
the errors on the power spectrum.  One would not want to go much
deeper over a smaller patch of sky than we consider, because we are
already masking roughly $10 \%$ of each data cube.  Without using the
increased sensitivity to remove more of the bright bad lines, going
deeper and shallower would increase the noise in the power spectrum
due to increased sample variance.

If the mask were not dependent on the integration time (e.g. obtained
from a different survey of foreground galaxies) it is straight forward
to determine in a given time what the optimal sky coverage is for
measuring power on a particular scale.  Minimizing Eq.~(\ref{ccerror})
with respect to time integrated per pointing, holding the total
observation time fixed, one finds that the optimal coverage sets
$P_{\rm noise1}P_{\rm noise2}=P_{1,2}^2+(P_{\rm 1total}-P_{\rm noise1})(P_{\rm 2total}-P_{\rm noise2})$.
The product of the detector noise power spectra equals the sum of the
sample variance contribution to the power spectrum uncertainty.

\begin{figure*}
\includegraphics[width=5in]{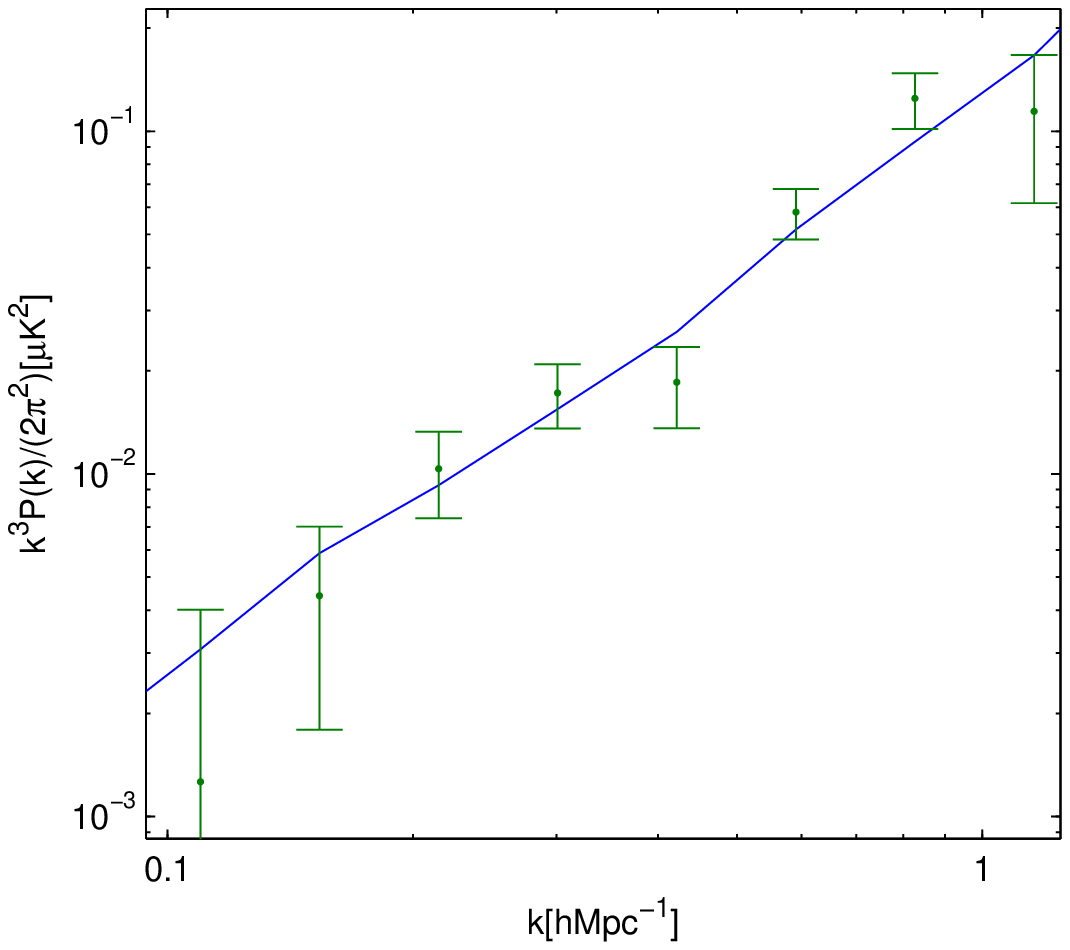}
\caption{\label{CO} The cross power spectrum of CO(1-0) and CO(2-1)
from a central redshift of $z=7.5$ measured with the telescope
described in $\S$4.  An integration of $3 \times 10^7$s and a redshift
range of $\Delta z = 0.9$ are assumed. The solid blue line is the
power spectrum of the CO line emission alone measured from our
simulated data.  The green points are the measurements of the power
spectrum recovered when the full simulated data set is used.  This
includes detector noise and bad line emission from the other lines in
Table \ref{tab}.  The error bars are calculated from
Eq.~(\ref{ccerror}), where we have estimated $P_{\rm1total}$ and
$P_{\rm2total}$ using our simulated data.  These errors include
detector noise, bad line emission and sample variance.}
\end{figure*}

\begin{table*}
\caption{\label{COtable} Summary of the results from the example of cross
correlating CO(1-0) and CO(2-1).  The RMS detector noise is the value
in each \emph{voxel}.  The bad line power to detector noise power
ratio gives the relative contributions to the statistical error in the
cross power spectrum due to the auto-correlations from all the bad
lines and the detector noise which appear in Eq.~(\ref{ccerror}).}
\begin{tabular}{l l l}
 & CO(1-0) & CO(2-1) \\
\hline 
Average Line Signals ($\bar{S}_{\rm line}$)& 0.1$\mu$K & 0.094$\mu$K\\ 
Fraction of Voxels Masked & 0.0 & 0.015  \\ 
RMS Detector Noise & 1.0$\mu$K & 0.7$\mu$K \\
Bad line Power$/$Noise Power ($k=0.1h^{-1}$Mpc) & 0.0  & 7.0  \\
Bad line Power$/$Noise Power ($k=0.3h^{-1}$Mpc) & 0.0 & 1.5 \\
Bad line Power$/$Noise Power ($k=0.8h^{-1}$Mpc) & 0.0 & 0.5 \\
Cross Power S$/$N per $k$-mode ($k=0.1h^{-1}$Mpc) & $\; \; \; \; \;$ 0.24   \\
Cross Power S$/$N per $k$-mode ($k=0.3h^{-1}$Mpc) & $\; \; \; \; \;$ 0.12  \\
Cross Power S$/$N per $k$-mode ($k=0.8h^{-1}$Mpc) & $\; \; \; \; \;$ 0.05  \\
\hline
\end{tabular}
\end{table*}

\section{Intensity mapping CO(1-0) and CO(2-1)}
As another example we consider intensity mapping the cross
correlations between CO(1-0) and CO(2-1) at high redshifts with a
dedicated instrument currently being planned (J. Bowman 2011, private
communication).  Other similar instruments are currently being planned
(G. Bower 2011, private communication).  This observation consists of
two telescopes: a 20 meter dish and a 10 meter dish to observe CO(1-0)
and CO(2-1) respectively.  Each of these telescopes can simultaneously
observe 3 ${\rm deg^2}$ of the sky with angular resolution set by the
beam size (3.5-5 arcmin at $z=7-10$).  We assume a spectral resolution
of $R=(\nu / \Delta \nu) =1000$.  While the actual instrument will
have a higher resolution this is sufficient to measure fluctuations on
the scales we consider.  To determine the detector noise we use the
radiometer equation \cite{radiobook},
\begin{equation}
\sigma_T = \frac{T_{\rm sys}}{\sqrt{2t \Delta \nu}},
\end{equation}
where $T_{\rm sys}$ is the system temperature which we have assumed to
be 30K, $t$ is the integration time which we have assumed is $3 \times
10^7$s, and the factor of $\sqrt{2}$ appears in the denominator
because the intensity will be mapped from dual polarization.  We
create a synthetic data set for this instrument centered at $z=7.5$
and the recovered cross power spectrum is shown in Figure \ref{CO}.
We summarize some other properties of this simulated measurement in
Table \ref{COtable}.

\section{Discussion and conclusions} 
By cross correlating emission in different spectral lines from the
same galaxies, it is possible to measure their clustering.  This
clustering, quantified by the line cross power spectrum, can be
measured for galaxies which are too faint to detect individually, but
which can be observed in aggregate due to their large numbers
\cite{VL}.

In this paper, we have shown that the line cross power spectrum can be
accurately measured with future instruments, based on synthetic data
created using cosmological dark matter simulations.  We produced our
synthetic data by associating dark matter halos with galaxies and
assigning each a spectrum.  The continuum was generated by scaling
that of M82 with the SFR in each halo and line emission was set by
calibrating with lower redshift galaxies.  The SFR was computed for
halos with an abundance matching technique calibrated to observations
of galaxy UV luminosity functions.  Our synthetic data also included
detector noise and bright emission due to astrophysical foregrounds
and backgrounds such as that from dust in our galaxy and the CMB.
Even if our simple prescription deviates somewhat from reality, it
still illustrates our main point, that whatever the underlying power
spectrum of emission from galaxies is, it can be measured with the
accuracy predicted analytically by Eq.~(\ref{ccerror}).  It is
reassuring that the complications addressed in our simulations such as
removal of bright astrophysical foregrounds and masking out bright bad
line emission do not hinder measurement of the power spectrum compared
to the analytic expectations from Ref.~\cite{VL}.

Measuring the line cross power spectrum consists of three main steps.
First, a smooth function such as a polynomial is fit to each \emph{pixel} on
the sky and subtracted from the data to remove smooth foregrounds and
the continuum emission from galaxies.  Next, bright \emph{voxels} are masked
out.  One must be careful in the masking technique as it is possible
to introduce spurious correlations if the masks are correlated with the
target lines which appear in both data cubes.  This can be avoided if
bright sources are found individually at high significance and the
corresponding \emph{voxels} with bright contaminating lines are set to zero.
Finally, the data is Fourier transformed and then the power spectrum
is averaged in spherical shells.  Modes corresponding to long
wavelengths along the line of sight are not included, because they are
contaminated during the fitting and subtraction step.

We find that the line cross power spectrum can be measured with the
accuracy predicted analytically by Eq.~(\ref{ccerror}), derived in
Ref.~\cite{VL}.  In particular, we tested two hypothetical instruments, an
instrument mounted on SPICA and a pair of large ground based
telescopes designed to measure the emission of CO(1-0) and CO(2-1).
Though not included in our examples, it would also be valuable to
measure emission from more than two different lines from the same
redshifts.  This could improve statistics and allow determination of the
ratio of line emission in different lines by taking the ratio of
different cross power spectra.

Our results suggest that cross correlating galaxy line emission is a
promising technique for studying high redshift galaxies.  It will
enable one to measure the evolution of the total line signal from all
galaxies at a particular redshift, even those that are too faint to be
resolved individually.  This could reveal details about the evolution
of galaxies' properties such as SFR density or average
metallicity.  It may also be possible to use these observations to
study the history of cosmic reionization, both by estimating the
ionizing flux from faint galaxies and by looking for a sharp change in
signal versus redshift due to the change in the minimum mass of halos
which host galaxies.

\section{Acknowledgments}
We thank Judd Bowman for helpful discussions. This work was supported in
part by NSF grant AST-0907890 and NASA grants NNA09DB30A and NNX08AL43G
(for A.L.).

\bibliography{paper_avihy}

\end{document}